# Deciphering the Language of Nature: A transformer-based language model for deleterious mutations in proteins

Theodore T Jiang[1,2,3], Li Fang[1,4*], Kai Wang[1,5*]

[1] Raymond G. Perelman Center for Cellular and Molecular Therapeutics, Children's Hospital of Philadelphia, Philadelphia, PA 19104, USA

[2] Palisades Charter High School, Pacific Palisades, CA 90272, USA

[3] Massachusetts Institute of Technology, Cambridge, MA 02139, USA

[4] Department of Genetics and Biomedical Informatics, Zhongshan School of Medicine, Sun Yat-sen University, Guangzhou 510080, China

[5] Department of Pathology and Laboratory Medicine, Perelman School of Medicine, University of Pennsylvania, Philadelphia, PA 19104, USA

*Correspondence should be addressed to wangk@chop.edu (K.W.) and fangli9@mail.sysu.edu.cn (L.F.).

# Abstract

Various machine-learning models, including deep neural network models, have already been developed to predict deleteriousness of missense (non-synonymous) mutations. Potential improvements to the current state of the art, however, may still benefit from a fresh look at the biological problem using more sophisticated self-adaptive machine-learning approaches. Recent advances in the natural language processing field show transformer models–a type of deep neural network–to be particularly powerful at modeling sequence information with context dependence. In this study, we introduce MutFormer, a transformer-based model for the prediction of deleterious missense mutations, which uses reference and mutated protein sequences from the human genome as the primary features. MutFormer takes advantage of a combination of self-attention layers and convolutional layers to learn both long-range and short-range dependencies between amino acid mutations in a protein sequence. In this study, we first pre-trained MutFormer on reference protein sequences and mutated protein sequences resulting from common genetic variants observed in human populations. We next examined different fine-tuning methods to successfully apply the model to deleteriousness prediction of missense mutations. Finally, we evaluated MutFormer's performance on multiple testing data sets. We found that MutFormer showed similar or improved performance over a variety of existing tools, including those that used conventional machine-learning approaches. We conclude that MutFormer successfully considers sequence features that are not explored in previous studies and could potentially complement existing computational predictions or empirically generated functional scores to improve our understanding of disease variants.

## Introduction

Whole-exome and whole-genome sequencing technologies are powerful tools for the detection of genetic mutations. A typical human genome has 4.1 million to 5.0 million variants when compared with the reference genome sequence (The 1000 Genomes Project Consortium, 2015), while the average exome captures genomic regions that account for 1-2% of the human genome. Therefore, distinguishing or prioritizing a small number of disease-related variants from such a large number of background variants becomes a key challenge in understanding genome and exome sequencing data. In particular, interpretation of non-synonymous single nucleotide variants (nsSNVs) is of major interest, because missense mutations in proteins account for over half of the current known variants responsible for human-inherited disorders, especially Mendelian diseases, where the mutations have high penetrance (Stenson, et al., 2020). Unlike frameshift indels or splicing mutations in canonical splice sites that have high likelihood to alter protein function, missense mutations change only a single amino acid, thus, most of them may not have significant impacts on protein function. To this end, population-specific allele frequencies, such as those inferred from the ExAC (Karczewski, et al., 2017) and gnomAD (Karczewski, et al., 2020) databases, can be useful to filter out common missense variants that likely to be neutral, and mutation databases such as ClinVar (Landrum, et al., 2017) and HGMD (Stenson, et al., 2020) can be valuable resources to find previously reported mutations that may be deleterious. Still, a large number of missense variants from exome sequencing are not yet definitively documented; therefore, the functional interpretation of such variants remains a crucial task.

Numerous computational tools have been developed to predict the deleteriousness or pathogenicity of missense mutations (Liu, et al., 2011; Liu, et al., 2020; Liu, et al., 2016; Thusberg, et al., 2011), however, as shown by multiple recent publications, the accuracy of predictive algorithms still has room for improvement. Databases such as dbNSFP have now documented these whole-exome prediction scores for different prediction algorithms, in an effort to facilitate the development of improved functional assessment algorithms. However, depending on the evaluation data sets that were used, most algorithms for missense variant prediction are 65-80% accurate when examining known disease variants, and only approximately 43.4% of pairwise prediction correlations between different predictive algorithms are greater than 0.5 (Liu, et al., 2020). Many conflicting predictions can be made between different algorithms, which motivated the development of several "ensemble"-based scoring systems that combine multiple prediction algorithms, such as MetaSVM, REVEL, CADD and others. In fact, predictions combined from different algorithms are considered as a single piece of evidence according to the ACMG-AMP 2015 guidelines (Richards, et al., 2015). In addition, most existing computational algorithms are based on similar or related information (e.g., evolutionary conservation scores, mutation tolerance scores); potential improvements to the current state of the art could benefit from a fresh look at the biological problem using more sophisticated self-adaptive machine-learning approaches that examine additional types of information.

In other previous prediction algorithms, deep learning-based sequence-focused models have been demonstrated as effective in modeling variant function. These existing methods primarily utilized convolutional neural networks (CNNs) to model sequences; however, recently, advances in deep learning have shown transformer models to be particularly powerful for modeling sequential data. The modern transformer model, Bidirectional Encoder Representations from Transformers (BERT) (Devlin, et al., 2019; Vaswani, et al., 2017) relies on its central mechanism, the self-attention. The usage of the self-attention allows the transformer model to achieve unprecedented ability to model relationships between tokens in a sequence – crucial in the comprehension of linear sequences. In the past three years, transformers have achieved state-of-the-art performances on a broad range of natural language processing (NLP) tasks (Devlin, et al., 2019; Lan, et al., 2019; Liu, et al., 2019; Yang, et al., 2019), and transformers are competitive with more traditional CNN-based models on image recognition tasks (Dosovitskiy, et al., 2020). As of late, transformers have also been successfully applied for modeling protein structure in Alphafold2 (Jumper, et al., 2021), and in works such as Enformer (Avsec, et al., 2021), which used transformers for DNA interpretation. Part of the reason for the success of transformers may be due to its increased ability to handle subtle context dependency through a multi-head attention mechanism, and the ability to compute attentions in parallel to greatly speed up computation over typical RNN-based algorithms.

In biological contexts, each amino acid in a given protein sequence exerts its function in a context dependent manner, including both local dependency (such as forming a short signal peptide that was recognized by cellular machinery) and long-distance dependency (such as being close to another amino acid in three-dimensional structure to form a binding site for ligands). Therefore, we hypothesize that transformer models would be capable of a more effective modeling of protein sequences, somewhat similar to how transformers have 'transformed' the field of natural language processing and language translation over the past few years.

In this study, we propose MutFormer, a transformer-based model to assess deleteriousness of missense mutations. MutFormer is an adaption of the BERT architecture (Devlin, et al., 2019) to protein contexts, with appropriate modifications to incorporate protein-specific characteristics. MutFormer can analyze protein sequences directly, with or without any homology information or additional data. Our experiments show that MutFormer is capable of matching or outperforming current methods in the deleteriousness prediction of missense variants.

MutFormer is based on the BERT architecture (Devlin, et al., 2019). A central component of the classical BERT model is its bidirectional self-attention. This mechanism uses a 2D matrix to model context between all positions in a given sequence, enabling efficient learning of long-range dependencies between residues. The convolution, on the other hand, another mechanism capable of learning dependencies, is better suited for short-range dependencies: convolutions are more capable of prioritizing localized patterns via filters, while the repeated application of convolution filters, which are required for the relating of farther residues in a sequence, often weakens long range dependencies. MutFormer takes advantage of both self-attention layers and convolutional layers to effectively learn both long-range and short-range dependencies.

In language modeling tasks, words or sub-words are short-range features of the sequence. The original BERT uses a fixed WordPiece vocabulary which contains common words/sub-words in the training corpus (Wu, et al., 2016). This vocabulary cannot be tuned during the pre-training and fine-tuning process; therefore, a spelling error may introduce an out-of-vocabulary word that will hinder the model's interpretation ability of a given sequence. In protein sequences, "words" correspond to key subsequences or patterns of amino acids. These "words" can often be changed due to mutations, and furthermore, are unknown. Recent studies showed that vocabulary-free models (e.g. byte-level models) are more robust to noise and perform better on tasks that are sensitive to spelling (Xue, et al., 2021). Therefore, instead of using a fixed vocabulary, convolutional layers placed in between the embedding layers and the transformer body are used by MutFormer. MutFormer uses these convolutions to learn its own "vocabulary" over the course of the training process, incorporating nonlinear patterns via the convolution filters (Figure 1). The weights of the convolutional layers are tuned during both the pre-training and fine-tuning processes.

# Methods

## The MutFormer model

The MutFormer architecture was implemented on top of the classic BERT architecture. The MutFormer model consists of 3 primary parts: embeddings, convolutions, and transformer body.

The embedding layers create position, label, and token embeddings. For MutFormer, position embeddings are calculated based on a learnable parameter matrix of size (max sequence length, embedding depth). Depending on the actual length of the sequence, the position embedding of that sequence is this parameter matrix sliced from the beginning such that its final shape is (sequence length, embedding depth). Label embeddings for MutFormer are calculated based on a (2, embedding size) learnable matrix. The label embeddings are used during deleteriousness prediction, specifically the pair method, which uses label identities of 0 or 1, indicating whether the residues belong to the reference or mutated sequence. The input label sequence is a 2D (sequence length x 2) array with one label for each residue represented one-hot. During pretraining and when utilizing deleteriousness prediction methods other than pair method, the label embeddings are not used. MutFormer's token embeddings embed the amino acid ids in each sequence into an embedding space. Token embeddings are calculated using a fixed learnable parameter matrix of size (vocab size, embedding size). The input tokens are a 2D matrix of size (sequence length, vocab size), where each residue is represented one-hot format. This input token matrix is matrix multiplied with the parameter matrix to get the embedding representation of the input sequence residues. The final embedding output is the sum of these thee embedding representations with layer normalization applied after the sum.

For the convolutions of MutFormer, two different implementations into the model were tested: 1) the above proposal from the Introduction section, where the convolution outputs are the only input to the following transformer body, and 2) an integrated approach with skip connections in which the original embedding outputs and convolution outputs are all fed into the transformer body as input (the embedding output is instead a sum of the embedding output and the result of applying two or four convolutions). Four convolutions were used in succession for each model. Each convolution used a kernel size of 3, and no pooling or concatenation was applied (only the convolution operation itself was necessary to take advantage of convolutions' pattern recognition ability).

The transformer body of MutFormer is taken from the original BERT model, along with its self-attention modules. Because both the position embeddings and self-attention modules have parameter matrices with limited size, a maximum input sequence length needs to be set. For each sequence that was longer than the maximum sequence length, the subsequence of interest is trimmed, and regardless of the trim position, the position embedding corresponding to position zero was always assigned to the first residue (for this reason, the position embeddings represent more of a relative position rather than true position within the overall protein sequence).

**Pretraining of MutFormer on human protein sequences**

We pretrained MutFormer on human reference protein sequences (all isoforms) and protein sequences caused by non-synonymous SNVs with > 1% population frequency in the gnomAD database (Karczewski, et al., 2020). The 1% threshold was used to ensure that this collection of examples contained a higher percentage of truly benign variants, since it is important for the model to learn the "syntax" of the language from "correct" examples during pretraining. The total number of protein sequences used during pretraining was 128,670. The maximum input length of MutFormer was set to 1024, where protein sequences longer than 1024 are cut into segments. Before cutting, a "B" letter was added to the beginning of a sequence and a "J" letter was added to the end of a sequence so that the true start and end of a protein sequence was also indicated ("B" and "J" are not included in the current biological amino acid alphabet). During cutting, all segments were retained, except for segments less than 50 amino acids long, which were discarded. In total, we accumulated 150,533 training datapoints from these 128,670 protein sequences, of which 86,213 were left intact when compared to their original protein sequence (with both "B" and "J" tokens), 23,857 contained a "B" token but not a "J" token, 21,672 contained a "J" token but not a "B" token, 18,791 contained neither a "B" or "J" token, and 2,561 original protein sequences did not have its tail end represented (because the tail end of the sequence was less than 50 amino acids long).

The original BERT model uses a self-supervised pretraining objective of recovering an original sequence from corrupted (masked) input, from which high-dimensional representations of the sequence are learned. Similar to BER, the pretraining objective of MutFormer was to predict corrupted amino acid residues from altered sequences. For the corrupted/masked residue prediction task, for each sequence, a number of residues were randomly selected for corruption. The ones that were selected were corrupted by either 1) replacing them with a [MASK] token or 2) a random other amino acid. This was done to encourage the learning of context not only around explicitly masked residues, but on the entire sequence. To ensure enough context was present for the model, a maximum number of 20 amino acids were masked per sequence. The maximum of 20 residues was chosen during testing; we tested training using a fixed masking percentage of 15%, however, this resulted in nonconvergence during pretraining. For this reason, the 20-residue cap was used alongside the 15% maximum, whichever one was lower, to determine the amount of residues masked. In order to facilitate the learning of more dependency knowledge and minimize overfitting, we utilized dynamic masking throughout the duration of training: the corrupted residues were changed randomly for each epoch of data trained on. Note that for MutFormer, the "next sentence prediction" objective used by the original BERT was removed, because protein sequences in aggregate, unlike their natural language counterpart, do not form 'paragraphs' with logical connections between sentences. The pre-training was performed on a single cloud Tensor Processing Unit (TPU) hardware accelerator (TPU v2-8) on the Google Cloud Platform.

We pretrained MutFormer (implementing convolutions) with three different model sizes (Table 1), as well as a MutFormer model with 8 transformer layers that utilized the integrated version of convolution implementation (Figure 1). For comparison purposes, we also pretrained two models without convolutions. We designated these models MutBERT, as indication of the usage of the original BERT architecture (Table 1). The hyperparameters used for pre-training are displayed in Table S1, and loss and accuracy for the pretraining task are listed in Table S2.

**Training and Evaluation data for variant deleteriousness prediction (Fine-tuning task)**

We obtained 84K manually annotated pathogenic missense SNVs from the Human Gene Mutation Database (HGMD) version 2016 (Stenson, et al., 2020). We combined this set with SNPs from the gnomAD database (Karczewski, et al., 2020) with allele frequency >0.1%, the vast majority of which are assumed to be benign. Although a commonly used allele frequency threshold for benign variants is 1%, we used 0.1% instead, in order to achieve 1) the exclusion of pretraining data from the fine-tuning data (MutFormer's pretraining data utilized the threshold of 1%); 2) the approximate balancing of the number of benign and deleterious/pathogenic examples in the dataset. Any mutations that appeared in both the HGMD database and benign set were removed, and from the remaining data, all the mutations present in the original pretraining data (i.e., >1% in gnomAD) were also eliminated. Within the training data, mutated sequences were obtained by mutating a reference protein sequence (using ANNOVAR) based on a nucleotide substitution specified by each variant in the dataset. In situations where the reference sequence residue did not match the reported reference residue by the mutation, the example was discarded. The final benign set contained 61K variants. The combination of the deleterious set and the benign set was randomly split into a training set and an independent validation set. The independent validation set was then isolated from the training set by reference sequence: to prevent memorization from the training set to the validation set, mutations were deleted from the training set if the mutated sequence or the mutation's reference sequence was present in the validation set. The independent validation set contains 5,282 benign variants and 3,145 deleterious variants. Note that this independent validation set, despite its independent selection, is still prone to bias because of its similarity with the training set. For this reason, this set is only used to internally compare the performance of MutFormer models trained within this study, and separate testing sets compiled from various different sources are used for the comparison of MutFormer with other existing methods of deleteriousness prediction.

**Fine-tuning MutFormer for prediction of deleterious mutations**

MutFormer was fine-tuned on the training data described above. Mutated protein sequences were generated using ANNOVAR (Wang, et al., 2010), with each sequence containing exactly one mutation. To obtain the best possible results in deleteriousness prediction, we tested three different fine-tuning methods (Figure 2):

1) Per residue classification uses a single mutated protein sequence as its input. The model is tasked with classifying each amino acid in the protein sequence as benign or deleterious. Amino acids that are identical to the reference sequence are labeled as benign, and the mutated residue is labeled as benign if the overall mutation is benign, or deleterious if the overall mutation is deleterious, depending on the true classification of the sequence (loss and metrics for classification are calculated on only the mutation site). This corresponds to the token classification task or named-entity recognition (NER) task in NLP.

2) Single sequence classification uses a single mutated protein sequence as its input. The model is tasked with classifying the entire sequence as deleterious or benign (via the [CLS] token). This corresponds to the sentence classification (e.g., sentiment analysis) in NLP.

3) Paired sequence classification uses both the reference sequence and mutated sequence as its input. The model classifies the aggregate of the sequences as deleterious or benign through comparison of the two sequences. This was inspired by the sentence similarity problem (e.g., the MRPC task) in NLP.

To find the best fine-tuning method, model, and sets of hyperparameters, we performed two different internal comparison tests using our independent validation set. Test 1 compared the MutFormer architecture with the classical BERT architecture, as well as the three different fine-tuning methods, using different hyperparameters. Test 2 compared the use of the integrated convolution implementation against the original MutFormer architecture. Prior to both tests, some initial testing (with our independent validation set) on random hyperparameters was done to establish a set of hyperparameter values that worked well with all combinations of methods/models included in the test. For test 1, the initial set of hyperparameters was established based on the three fine-tuning methods (per residue, single sequence, paired sequence) and three different models (MutBERT$_{8L}$, MutBERT$_{10L}$, MutFormer$_{8L}$) being tested. For test 2, the initial set of hyperparameters was established based on the four models (MutFormer$_{8L}$, MutFormer$_{10L}$, MutFormer$_{12L}$, MutFormer$_{8L}$ *(with integrated convs)*) being tested. Full list of hyperparameters used are detailed in Table 3, and results of test 1 and test 2 are displayed, respectively, in Figure 3 and in Figure 4, 5, S3.

To create a model capable of the best possible performance in deleteriousness prediction, in addition to utilizing protein sequence analysis, when training our final models (displayed in our final testing) (Figure 4), MutFormer also incorporated prediction values from previous methods published in literature. Previous methods' predictions were given to MutFormer as input in the following way. First, using ANNOVAR, predicted scores for all mutations within a newly generated test set were obtained from the dbNSFPv3 database (Liu, et al., 2016). These scores were standardized between 1 and 2, and all missing predictions were assigned values of 0. A fully connected dense layer was connected to these inputs, and the output of this dense layer was concatenated with the original model output. Another dense layer after this concatenated result was then connected to the output node to produce the end prediction result. This incorporation strategy prevents the model from becoming reliant on external predictions, limiting the weighting of sequence analysis vs external predictions to about 1:1 in MutFormer's prediction. This intended usage ratio is consistent to our findings when looking into the weights of the various finetuned models' prediction layers, where we found, for all models, an approximate weighting of half for MutFormer's sequence analysis, and half for external predictions.

**Testing MutFormer against existing methods of deleteriousness prediction**

To assess the performance of MutFormer against existing methods of deleteriousness prediction, a total of 5 testing datasets were used. Out of these 5 datasets, three are non-gene-specific and non-disease-specific datasets, and two are gene-specific mutation datasets (details for each testing dataset used are outlined in Table S2). For each dataset, filtering was performed using the reference sequences to ensure that no bias was present: all mutations that shared reference sequences with any mutation present in the pretraining data were deleted from the testing sets, and all mutants with identical reference sequences present in any of the independent test sets were removed from the fine-tuning training data prior to model training.

In order to allow for a more comprehensive evaluation of the performance of MutFormer with different levels of "fit" on a wide range of data (models with a higher "fit" will perform better on more similar data but worse on more dissimilar data; models with lower "fit" will have the opposite tendencies), different MutFormer models with varying hyperparameters that affected a model's level of "fit" were trained (an analysis of MutFormer's performance with varying levels of fit is analogous to an analysis of an ROC-type curve, where the performances of MutFormer on similar vs. dissimilar data is compared for different fit levels). In this test, the number of freezing layers and batch size were varied, while all other hyperparameters were set to the best ones found during our hyperparameter test 2 (see above). From these results, testing sets 3-5 showed more variation with different fit parameters than sets 1-2 did (results from all test runs are summarized in Figure S2). All models were then tested on all testing datasets, and the overall best performing model across all testing datasets was used to represent MutFormer in our comparison. Batch sizes of 16, 32, and 64 were tested in conjunction with freezing layer numbers of 0, 5, 6, and 8 (full hyperparameter description in Table 3). Freezing layer numbers are defined as the number of transformer body layers that were frozen (for MutFormer 8L with integrated convolutions, our current best performing model, 8 layers is the total number of transformer body layers). For any freezing layer number greater than 0, the embedding layers were frozen as well (through testing on our validation set we found that leaving the embedding layers trainable while freezing the transformer layers significantly decreased performance). Each model was trained for 14k steps, and checkpoint steps 6k, 8k, 11k, 12k, and 14k were evaluated.

For each dataset, MutFormer is shown twice: once without the incorporation of other scores and only relying on sequence data alone (labeled as "MutFormer (external predictions removed)" in Figure 4, 5 and S3), and another with the usage of external predictions as part of its input as described in the above "Fine-tuning MutFormer for prediction of deleterious mutations" section (labeled as "MutFormer" in Figure 4, 5 and S3). For MutFormer without incorporation of other scores, like MutFormer with the incorporation of external scores, we tested various models with varying levels of "fit," with an initial set of hyperparameters found through testing based on our independent validation set. Full set of hyperparameters for both of these MutFormer models are displayed in Table 3.

When fine-tuning our final models (used in our comparison of MutFormer vs existing methods), to increase MutFormer's overall generalization ability and limit overfitting, data augmentation was implemented for the fine-tuning training data: for all epochs of data, each datapoint had a 50% chance of being altered. Those that were selected to be altered would be trimmed down to anywhere from 50% length of the original sequence to 100% length of the original sequence. Trimming was done around the mutation site to ensure the mutation site on average stayed in the same location (in the middle) in the sequence before and after trimming. Epochs were also shuffled independently of each other.

## Results

### Effect of MutFormer's use of Convolutions on the pretraining task

During pretraining, we trained models of different sizes for both the MutFormer architecture and MutBERT (MutFormer model without convolutions) architecture (Table 1). The loss and accuracy on the pretraining task are shown in Table S2.. According to our results, the accuracy of MutFormer$_{8L}$ on the pretraining task test set was 54.5% higher than MutBERT$_{8L}$, indicating the advantage of the convolutions. In addition, the accuracy of MutFormer$_{8L}$ was 27.7% higher than that of MutBERT$_{10L}$, despite the latter having two more transformer layers and 10M more parameters (the subscript of each model name indicates only the number of transformer layers but does not consider the two convolutional layers). This outperformance despite smaller size verifies that the improved performance of the model with convolutions was not simply due to the additional number of parameters or additional layers.

### Performance of different fine-tuning methods and hyperparameters

As part of our internal comparison test 1, we fine-tuned MutFormer and MutBERT using three methods: per residue classification, single sequence classification and paired sequence classification (Figure 2, see Methods for details). The Receiver Operator Characteristic (ROC) curves and corresponding Area Under Curve (AUC) for deleteriousness prediction are shown in Figure 3A-C. Figure 3D shows a summary of the performance comparison of the three methods; paired sequence classification performed best, followed by per residue classification, and the optimal results were achieved by using a maximal input sequence length of 512 (usage of the paired sequence method means an aggregate sequence length of 1024) (Figure 3F). Upon examination of the performance of different model architectures, as shown in Figure 3E, MutFormer$_{8L}$ outperformed MutBERT$_{8L}$ and MutBERT$_{10L}$ for each fine-tuning method, indicating the advantage of the MutFormer architecture.

**Analyzing bias present in MutFormer's fine-tuning data**

To gauge the amount of bias present in the resulting fine-tuning dataset, distributions of positive and negative examples per protein reference id is displayed in Figure S4A. However, since certain proteins are naturally more sensitive to mutations than others, in order to appropriately interpret the positive and negative example distributions in the context of the human population, we graphed the probability of loss of function intolerance (pLI) score, obtained from GnomAD data for each corresponding protein ID in Figure S4B. From the two graphs, we see that protein IDs with relatively low pLI scores tend to have less deleterious representation and more benign representation in our fine-tuning dataset, suggesting that there is little bias introduced.

**MutFormer's Use of Integrated Convolutions**

In our internal comparison test 2, two different strategies for implementing the convolutions were tested: classic MutFormer, and MutFormer *(**with integrated convolutions**)*. Justification for the second implementation strategy is as follows: while the convolutions should in theory be able to create a representation of a protein that will enable the model to best interpret the sequence, some information that is present in the original raw embedded sequence may be lost in practice through the convolutions. To solve this, the integrated convolutions model, instead of feeding the embeddings through the convolutions linearly, utilizes skip connections that result in the convolutions acting as an "integrated" part of the original embedding layers, allowing the transformer model to access both the convolution filtered representation of the sequence as well as the original embedded representation. In our internal comparison test **2**, MutFormer *(with integrated convolutions)*'s performance in paired sequence classification of our independent validation set to that of the other three original MutFormer architecture models (Figure S1). ROC curves are shown in Figure S1A, and a summary comparison histogram of the four different models tested is shown in Figure S1B. Overall, margins of difference are small, but based on the results, the performance of the MutFormer model with integrated convolutions is higher than that of the original MutFormer model; even with only 8 transformer layers, the integrated convolutions model outperformed MutFormer$_{12L,}$ which had more layers and generally better prediction ability than MutFormer$_{8L}$.

**Comparison with existing variant prediction methods**

As paired sequence classification performed best, for the comparison of MutFormer versus other methods, this fine-tuning method was used. MutFormer's best overall performance was achieved by training our best model, MutFormer *(with integrated convs)*, on a batch size of 32, and 0 freezing layers (for full hyperparameter descriptions see Table 3). MutFormer's performance was compared against a variety of existing methods of deleteriousness prediction (Adzhubei, et al., 2010; Carter, et al., 2013; Choi, et al., 2012; Chun and Fay, 2009; Davydov, et al., 2010; Dong, et al., 2015; Garber, et al., 2009; Gulko, et al., 2015; Kircher, et al., 2014; Ng and Henikoff, 2003; Pollard, et al., 2010; Qi, et al., 2021; Quang, et al., 2015; Rentzsch, et al., 2019; Reva, et al., 2011; Schwarz, et al., 2014; Schwarz, et al., 2010; Shihab, et al., 2013; Shihab, et al., 2015; Siepel, et al., 2005; Wu, et al., 2021). In our comparison, existing methods' predictions were processed in the following way (note that this differs from the incorporation strategy of external scores into MutFormer as input): each method's predictions were standardized from 0-1, based on prediction values of all possible missense mutations present in the dbNSFPv3 database (Liu, et al., 2016). Missing predictions were automatically assigned a prediction value of 0. Both non-inverted and inverted prediction identities (1=deleterious, 0=benign and 0=deleterious, 1=benign) were tested across our fine-tuning training data, and inversion of scores was done accordingly for each algorithm being compared. We generated both an ROC curve, and due to the unbalanced nature of some of our datasets, a Precision Recall Gain curve for each dataset. For dataset 4, which only included rare benign examples, a threshold for each existing method, chosen by taking the point closest to the upper left corner on an ROC curve based on MutFormer's fine-tuning training data, was used to calculate a method specificity for each method. Upon analyzing the performances of the different testing datasets, we found that the best overall performing MutFormer model outperforms previous methods of deleteriousness prediction in non-gene-specific and non-disease-specific datasets (more similar to MutFormer's fine-tuning training dataset: sets 1-3On the two gene-specific databases (sets 4-5), which contain data less similar to that of MutFormer's fine-tuning data, MutFormer's performance in comparison to other methods expectedly drops, while still matching the performance of various existing methods. For MutFormer without external predictions, its performance is first among non-MutFormer methods for sets 2-3, among the top for set 1, and drops further than MutFormer with incorporated external predictions for sets 4-5. ROC curves for the full testing results are displayed in Figure 4, and precision recall gain curves for datasets 1, 2, 4, and 5 are displayed in Figure 5. Due to the large number of existing methods compared against, we also used the Delong test of ROC to statistically assess the pairwise probability that methods' ROC curve were identical. Delong test results for datasets 1, 2, 4, and 5 are displayed in Figure S3.

**MutFormer Complementing Evolutionary Approaches**

Due to MutFormer's primary reliance on protein sequence data analysis, we hypothesized that MutFormer would be capable of providing complementary information to existing evolutionarily based methods for variant classification. To assess this, we compared MutFormer (external prediction included) to EVE (Frazer, et al., 2021) using the provided scores and prediction data from EVE's database. Only examples with reported clinical significance of benign, pathogenic, likely pathogenic, or likely benign were used. We calculated the Spearman Rank Correlation between MutFormer and EVE for mutations predicted by both MutFormer EVE. We found rank correlations of 0.320 and 0. 336 (both with p values less than $1\times10^{-10}$) between MutFormer and EVE_ASM and between MutFormer and EVE_BPU, respectively. This rank correlation indicates that while MutFormer is correlated with EVE, the information provided by MutFormer is not identical to that which EVE provides, suggesting that MutFormer's ability to analyze sequences provides additional information to the protein interpretation problem which evolutionary approaches may not be capable of extracting.

**ProteinGym Evaluation**

In the interest of further assessing MutFormer's ability in interpretation of protein sequences, we evaluated MutFormer's performance on Deep Mutational Scanning (DMS) data. Using the ProteinGym database provided by Tranception (Notin, et al., 2022), we compared (external predictions included) MutFormer's performance on this set against all other methods provided by the ProteinGym database. Because some amino acid substitutions present in ProteinGym are not obtainable through SNPs, these amino acid substitutions were not included in our evaluation. After eliminating these examples from all method's data, we used ProteinGym's evaluation script to run evaluation of all methods and Mutformer on the remaining DMS data. Full results of this evaluation are displayed in Table S4.

**Analyzing the weights of MutFormer**

The MutFormer model takes advantage of both convolutions and attention, both of which contain useful information about MutFormer's interpretation mechanisms for protein sequences. To better understand these mechanisms, we analyzed MutFormer's model weights for the final best performing MutFormer model. For our analysis of the convolution layers, two visualizations were generated. In our first visualization, we plotted the average convolution filter weights for each of the four convolution operations. In our second visualization, based on each of the four convolution operations, we plotted the overall percentage consideration given to each input residue id (amino acid identity/special tokens). Both of these visuals are displayed in Figure S5A. Based on visual 2, we see that the overall considerations placed on each residue are very nearly identical across the four different convolutions. Special tokens, as expected, were given very large weight in overall consideration. Interestingly, the "J" token was given relatively significant weighting, but the "B" token was not as relatively prioritized. Of note, amino acids Serine (S), Cysteine (C), and Valine (V), took more priority relative to other amino acids. Based on visual 1, we can see that first, there are different prioritizations of positions for each of the four convolutions, and additionally, that while specific patterns cannot be directly derived from these filters, it is reasonable to infer that combinations of residue IDs and residue ID patterns are learned by different convolutions.

For our analysis of attention weights, we opted for two case studies obtained from the PDB database: 1B1C and 1P4O. The chosen proteins were selected because of their relatively large number of deleterious mutations and complex structure. We ran MutFormer with each protein sequence as input and generated four different visuals: 1) all 8 individual attention maps for each individual attention layer, 2) a rollout attention map representing the log scale result of taking the dot product over all attention outputs, 3) the distance map between residues within the protein's 3D structure, with each point value given as the negative of the 3D distance between the two residue positions, 4) a deleterious mutation map taken from HGMD data displaying the locations of deleterious mutation sites. All four visuals for each of the two proteins are displayed in Figure S5B. We can see from these visuals that while MutFormer's attention maps for each example do not perfectly correspond to either the 3D distance map or deleterious mutation location map, there is still notable resemblance present. For instance, in 1B1C's rollout attention, we find resemblance with the deleteriousness map, as both have bright bands roughly at positions 12, 48, 74, 120, and 160. We also observe some resemblance between the rollout attention of MutFormer for 1P4O and the 3D distance map, we roughly observe a bright rectangle from positions 175 to the end of the sequence, a broad band from positions 150 to 200 running the length and width of the rectangle, and a rough rectangle from positions 0 to 100.

**Precomputed deleteriousness scores for all missense mutations**

To facilitate future use by other studies, we precomputed deleteriousness scores for all missense mutations using the best performing MutFormer model. The inference was done on a cloud TPU device (v2-8), which took about 11.5 hours. These scores can be directly used in the ANNOVAR software to annotate missense variants identified from genome or exome sequencing, and being organized in a flat file format, can be also used in other functional annotation software tools to complement the dbNSFP database, which has a variety of other prediction scores for missense mutations in the human genome.

## Discussion

In the current study, we present MutFormer, a transformer-based machine-learning model to predict the deleteriousness of non-synonymous SNVs using protein sequence as the primary feature. We pretrained MutFormer on reference protein sequences and alternative protein sequences resulting from common genetic variants in the human genome and tested different fine-tuning methods for deleteriousness prediction. During our evaluation processes, MutFormer outperformed multiple commonly used methods and had comparable performance with other methods even when tested on data sets that were less similar to MutFormer's training data (gene-specific data, sets 5-6). Below we discuss several advantages and limitations of the MutFormer method and its computational package.

Although a large number of computational tools have been developed over the years on predicting the deleteriousness of non-synonymous mutations, to the best of our knowledge, MutFormer is among the first batch of tools that utilize transformer models to adapt the biological sequence analysis problem as a language analysis problem. A similar model of note is ProtBERT (Elnaggar, et al., 2021), a previous application of the BERT architecture to protein contexts. Despite both utilizing the transformer architecture, MutFormer differs from ProtBERT in several key aspects: 1) MutFormer makes use of convolutions to learn its own "vocabulary" while ProtBERT uses a fixed vocabulary; 2) MutFormer was pretrained on human protein sequences and common variants, while ProtBERT was trained on reference sequences of all species with sequence information; 3)

MutFormer's primary goal was deleteriousness prediction while ProtBERT focused on subcellular localization of proteins and secondary structure prediction.

In addition, the training process of MutFormer is simple and straightforward. MutFormer utilizes a self-supervised pretraining strategy and therefore does not require any labeled data. For this reason, a large model with hundreds of millions of parameters can be trained on a large amount of non-curated data. On this note, while the current study focused on human genome exclusively, it is conceivable to include other well-annotated genomes from other species in future studies to see whether increased complexity in the sequence space during pre-training can further improve performance. In the fine-tuning stage, MutFormer learns the deleteriousness of mutations based on the labeled training data as well as its understanding of protein sequence already learned in the pretraining stage, allowing a small amount of fine-tuning data to be effectively used to achieve an accurate result. Furthermore, transformers consider attention, which is not only useful for understanding context in language processing problems, but could also give important insights into the deleteriousness effect amino acids can have under different sequence contexts. Even though direct feature attribution is not possible with a model containing millions of intertwined model parameters such as MutFormer, our analyses of MutFormer's model weights show that MutFormer is able to prioritize and model these relationships through the use of convolutions and attention.

There are also several limitations of the current study. First, the training data and testing data sets are still of limited size, and testing on large-scale experimentally or clinically supported datasets would result in a more effective evaluation of usability. In the future, MutFormer can be evaluated on large-scale genome sequencing data followed by manual review, to determine whether it helps prioritize deleterious variants in clinical sequencing settings. Second, because of computational limitations, we did not fully test all parameters during training. As a result, it is likely that our results are not completely optimized; larger models using longer maximum sequence lengths would also be able to outperform the current MutFormer models (for example, a 12L MutFormer with integrated convolutions should perform noticeably better than the best current MutFormer model with only 8 attention layers). Third, in the deleteriousness prediction of missense mutations, it is likely impossible for a given model to obtain all required evidence from sequence data alone, so incorporation of other features, such as 3D structure (i.e., analyzing 3D structure to scale attention with 3D distance, or labelling sections as belonging to beta sheets or alpha helices for better prediction of deleteriousness), methylation, clinical phenotypic information (i.e., using this knowledge to prioritize certain genes), and other features that could significantly affect proteins' behavior, could reasonably improve overall understanding and thus performance.

In summary, MutFormer is a novel transformer-based method to predict the functional effects of missense mutations. We hope that MutFormer can bring new insights to the bioinformatics community, by being a language model capable of improving our understanding of the language of proteins. Given that MutFormer used complementary information that other bioinformatics tools developed for deleteriousness prediction, we also envision that they could be combined to reach consensuses on predictions, which may be useful in implementation into current clinical guidelines.

## Data and Code Availability

The source code to run MutFormer, all 6 pretrained models, a reproducible workflow, and the best performing fine-tuned models are available at the GitHub repository: https://github.com/WGLab/MutFormer/.

## Competing interests

## Acknowledgements

We would like to acknowledge the TPU Research Cloud (TRC) program by Google, which provided us with TPUs for the majority of the project. The study is in part supported by NIH grant GM132713 (KW) and the CHOP Research Institute.

# Tables

## Table 1

Model sizes of the pretrained models (subscripts in the models' names denote the number of self-attention layers in the respective transformer bodies) ("-" indicates same as above).

| Model Name | Hidden Layers | Hidden Size | # of parameters |
|---|---|---|---|
| MutBERT$_{8L}$ | 8 | 768 | 58M |
| MutBERT$_{10L}$ | 10 | 770 | 72M |
| MutFormer$_{8L}$ | 8 | 768 | 62M |
| MutFormer$_{10L}$ | 10 | 770 | 76M |
| MutFormer$_{12L}$ | 12 | 768 | 86M |
| MutFormer$_{8L}$ *(with integrated convs)* | 8 | 768 | 64M |

Note: MutFormer$_{12L}$ has the same size as BERT$_{Base}$

Additional hyperparameters constant for all runs:
- Intermediate Size: 3072
- Maximum Input Sequence Length: 1024

# Table 2

Details for each testing dataset:

| Testing dataset | Testing dataset composition | Compilation year |
|---|---|---|
| Set 1: Meta_SVM_LR_set_1 | Dataset compiled by a previous method, Meta_SVM/Meta_LR (Dong, et al., 2015). Used to assess Meta_SVM and Meta_LR's performance against other methods.<br><br>Composition after filtering:<br>• 56 pathogenic examples compiled from recent Nature Genetics publications at the time<br>• 35 benign examples from the CHARGE (Cohorts for Heart and Aging Research in Genetic Epidemiology) database, which focuses on identifying genes underlying heart, lung, and blood diseases. | 2015 |
| Set 2: Meta_SVM_LR_set_2 | Dataset from same source as set 2 (Meta_SVM_LR_set_1)<br><br>Composition after filtering:<br>• 4135 pathogenic examples from Varibench testing dataset II for missense mutations (Sasidharan Nair and Vihinen, 2013) (Varibench is a dataset designed specifically for the testing of prediction methods for pathogenicity).<br>• 5884 benign examples also from Varibench testing dataset II | 2015 |
| Set 3: Meta_SVM_LR_set_3 | Dataset from same source as set 2 (Meta_SVM_LR_set_1) set 3 (Meta_SVM_LR_set_2).<br><br>Composition after filtering:<br>• 2422 benign examples from the CHARGE database. | 2015 |
| Set 4: Varibench_PPARG | Dataset from Varibench (Sasidharan Nair and Vihinen, 2013). Compiled by a study that specifically aimed to create datasets for assessing computational models' performance in pathogenicity prediction of missense mutations (Li, et al., 2018). Focused on the PPARG gene which codes for the gamma member of the PPAR (Peroxisome Proliferator-activated Receptor) family of nuclear receptors, which can be linked to the pathology of diseases including diabetes, atherosclerosis, and cancer.<br><br>Composition after filtering:<br>• 145 pathogenic variants from the experimentally validated Missense InTerpretation by Experimental Response (MITER) database.<br>• 2207 benign variants from the same source | 2018 |

| Set 5: Varibench_TP53 | Dataset from same source as Set 5 (Varibench_PPARG). Focused on the TP53 gene which codes for tumor protein p53, a tumor suppressor gene.<br><br>Composition after filtering:<br>• 608 pathogenic examples from the IARC database (database specific for TP53), labeled for significantly changing the gene expression level of the TP53 gene.<br>• 531 benign examples also from IARC which did not change expression level significantly. | 2018 |
|---|---|---|

# Table 3

MutFormer Fine-tuning hyperparameter specifications (1 of 2):

| Test / model | Model Architecture | Fine-tuning method | Initial / end Learning Rate | Training steps |
|---|---|---|---|---|
| Internal comparison test 1 | (MutFormer$_{8L}$, MutBERT$_{8L}$, MutBERT$_{10L}$) | (Per residue, Single Sequence, Paired Sequence) | 1e-5 / (2e-7 to 1e-6) | (4k (fine-tuning method 1 and 2), 10k (fine-tuning method 3)) |
| Internal comparison test 2 | (MutFormer$_{8L}$, MutFormer$_{10L}$, MutFormer$_{12L}$, MutFormer$_{8L}$ *(with integrated convs)*) | Paired Sequence | 1e-5 / 1.4e-6 | 12k |
| MutFormer Comparison with others | MutFormer$_{8L}$ *(with integrated convs)* | Paired Sequence | 1e-5 / 3e-9 | 14k (evaluated on 6k, 8k, 11k, and 12k) |
| MutFormer final model (with external predictions) | MutFormer$_{8L}$ *(with integrated convs)* | Paired Sequence | 1e-5 / 3e-9 | 12k |
| MutFormer final model (no external predictions) | MutFormer$_{8L}$ *(with integrated convs)* | Paired Sequence | 1e-5 / 3e-9 | 8k |

MutFormer Fine-tuning hyperparameter specifications (2 of 2):

| Test / model | Max Input Sequence Length | Batch Size | Weight Decay | Freezing layers | External predictions? |
|---|---|---|---|---|---|
| Internal comparison test 1 | (64, 128, 256, 512) | 16 | 0.01 | 0 | No |
| Internal comparison test 2 | (256, 512) | (16, 32, 64) | 0.01 | 0 | No |
| MutFormer Comparison with others | 512 | (16, 32, 64) | 0.01 | (0, 5, 6, 8) | Yes |
| MutFormer final model (with external predictions) | 512 | 32 | 0.01 | 0 | Yes |
| MutFormer final model (no external predictions) | 512 | 32 | 0 | 0 | No |

Additional hyperparameters constant for all runs:
Gradient Clip amount: No gradient clipping was used during fine-tuning (this was used previously during pre-training)

## Figure legends

### Figure 1

**The MutFormer model architecture.** A general outline of MutFormer's model architecture. A system of positional, label, and token embeddings is used to first vectorize the input tokens. Four convolution layers are used to process the embedding representation (for the integrated convolutions model, skip connections are used). A bidirectional transformer body with self-attention applies a sequence of attention layers to the resulting embedding representation to obtain the output embeddings. The output embeddings are used for token or sequence level classification.

### Figure 2

**Different fine-tuning methods tested in this study**. **A**) Per residue classification. The input is a protein sequence that contains exactly one variant. Each residue (amino acid) is given a label of benign/deleterious. Benign variants and residues that are identical to the reference sequence are labeled as benign. The fine-tuning task is to predict the label of each amino acid. This is similar to token classification problems (e.g., named-entity recognition) in NLP. **B**) Single sequence classification. The input is a protein sequence that contains exactly one variant with unknown significance. The embedding of the [CLS] token in the last layer is used to predict whether the sequence contains a deleterious variant. This is similar to sentence classification problems (e.g., sentiment analysis) in NLP. **C**) Sequence pair classification. The input is a pair of two sequences: a reference protein sequence and a mutated protein sequence (with a benign or deleterious variant in the center). The embedding of the [CLS] token in the last layer is used to predict whether the mutated sequence contains a deleterious variant. This is similar to sentence pair classification problems (e.g., sentence similarity) in NLP.

**Figure 3**

**Fine-tuning internal comparison test 1: Performance comarison of different fine-tuning methods and MutFormer architecture/MutBERT architecture. A-C)** ROC curves for two different model architectures (MutFormer (without integrated convolutions)/MutBERT) and three fine-tuning methods (per residue, single sequence, paired sequence). Labels are in the following format: "mn_{model name}_sl_{max input sequence length}: {ROCAUC}". **D**) Performance comparison of three different fine-tuning methods, using AUC scores shown in panels A-C. Whiskers indicate min and max values. **E**) Performance comparison of three pretrained models: MutBERT$_{8L}$, MutBERT$_{10L}$ and MutFormer$_{8L}$. **F**) Performance of different max input sequence lengths. **E, F**) The results are mean ± Standard Error of the Mean (SEM).

**Figure 4**

**ROC Curves for Performance comparison with existing methods.** ROC curves and accuracy / AUC scores of MutFormer and different existing methods of deleteriousness prediction evaluated on 5 different databases. Labels are formatted in the following way: {Method}: ROC={ROC AUC} **A:** Meta_SVM_LR_set_1 – dataset compiled by a previous paper that originally outlined the MetaSVM and MetaLR methods, containing 56 negative examples and 35 positive examples. **B:** MetaSVM_LR_set_2 – same source as Meta_SVM_LR_set_1, containing 5866 negative examples and 4115 positive examples. **C:** MetaSVM_LR_set_3 – same source as Meta_SVM_LR_set_1 and set_2, containing 2422 negative examples (because only negative examples are present, ROC is invalid in this case; instead, specificity is used for comparison). **D:**

Varibench_PPARG – dataset from Varibench for the peroxisome proliferator activated receptor (gamma) gene, containing 4671 negative examples and 3428 positive examples. **E:** Varibench_TP53 – dataset from Varibench for the TP53 gene which codes for the tumor suppressor P53 protein, containing 3444 negative examples and 4505 positive examples.

**Figure 5**

**Precision Recall Gain Curves for Performance comparison with existing methods.** Precision Recall Gain Curves / AUC scores of MutFormer and different existing methods of deleteriousness prediction. Labels are formatted in the following way: {Method}: AUC={Precision Recall Gain AUC}. **A:** Meta_SVM_LR_set, **B:** MetaSVM_LR_set, **C:** Varibench_PPARG, **D:** Varibench_TP53.

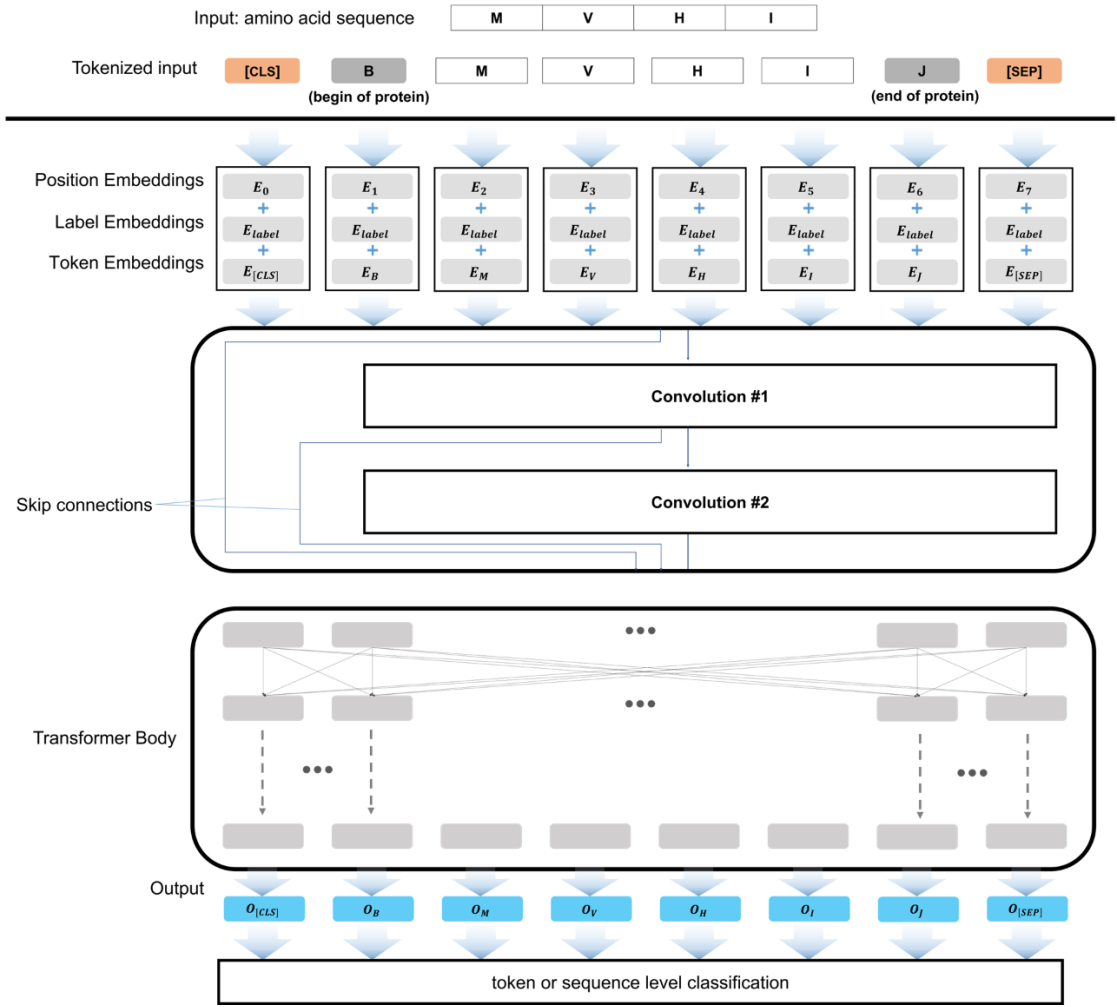

**Figure 1**

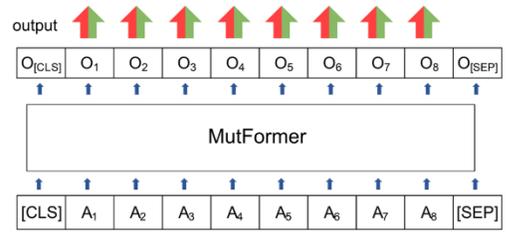
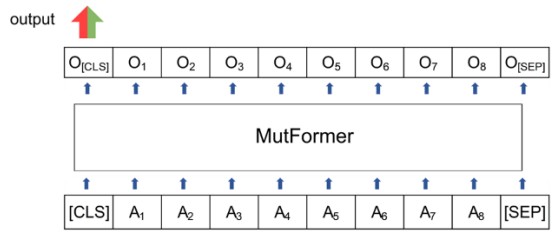
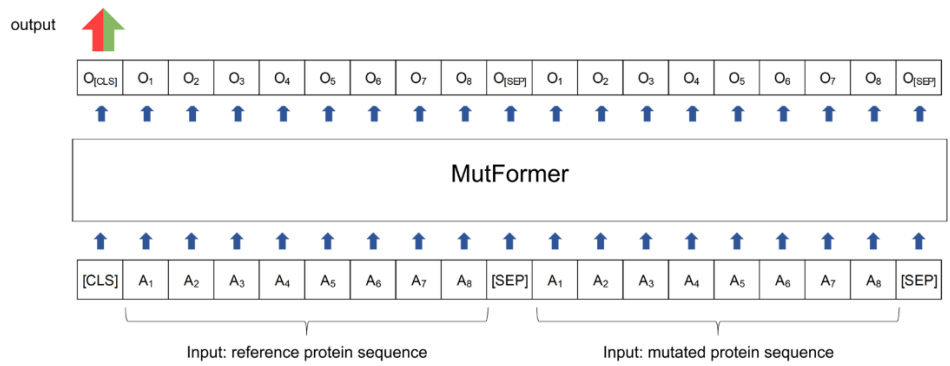

**Figure 2**

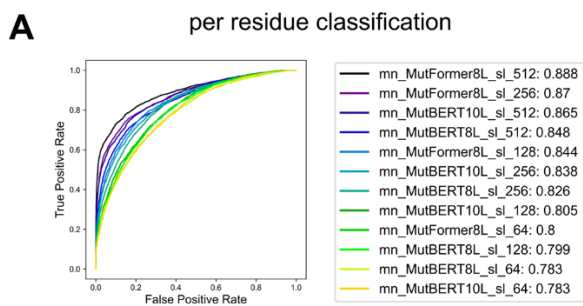
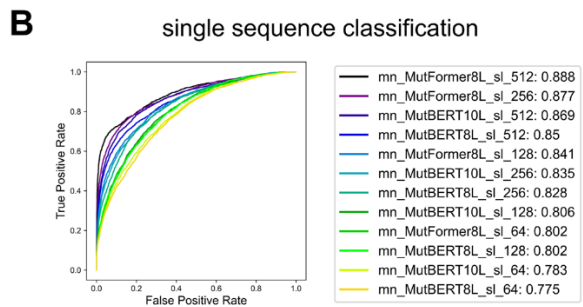
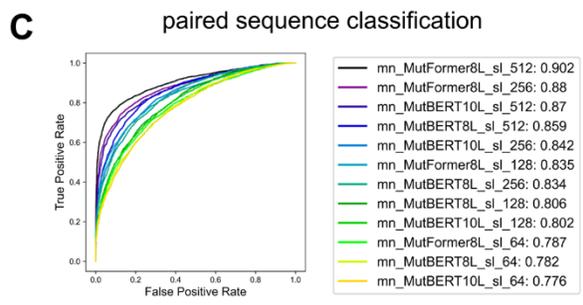
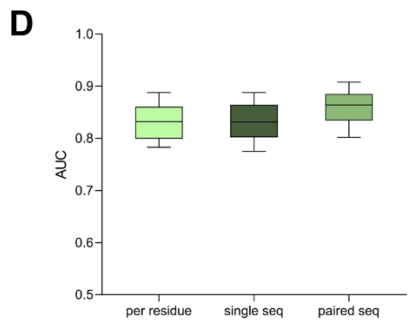
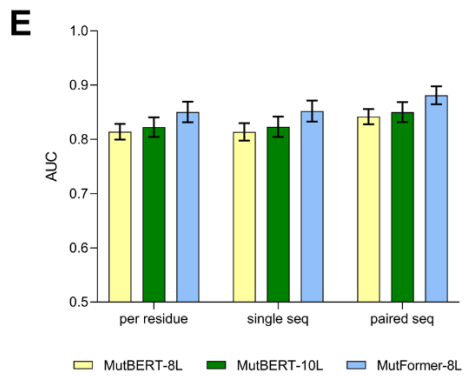
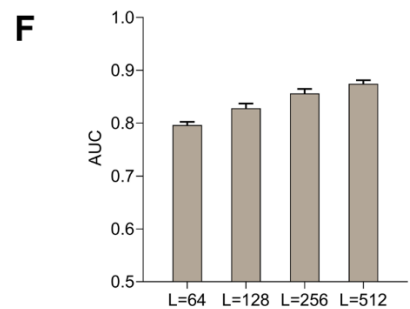

**Figure 3**

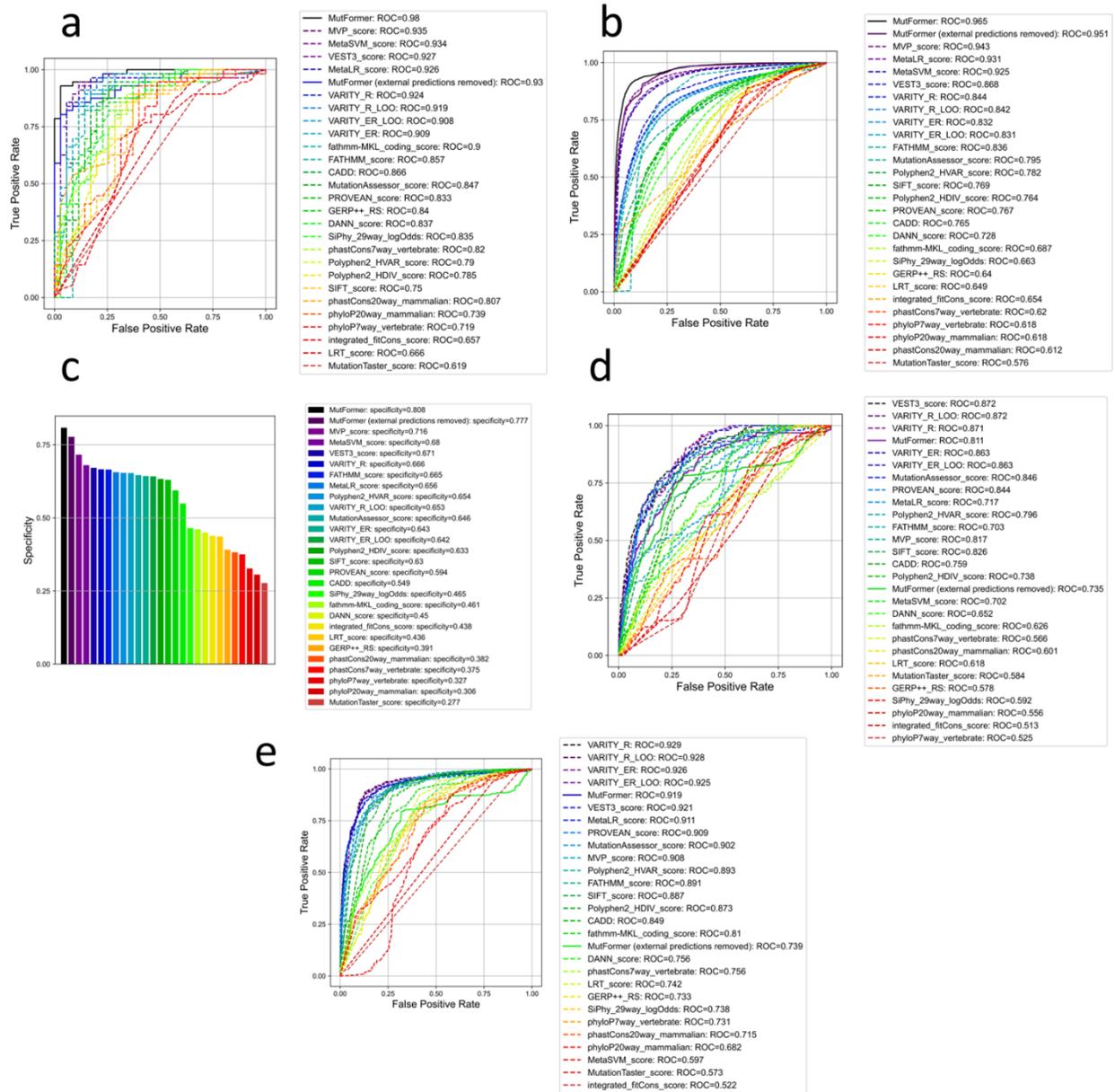

**Figure 4**

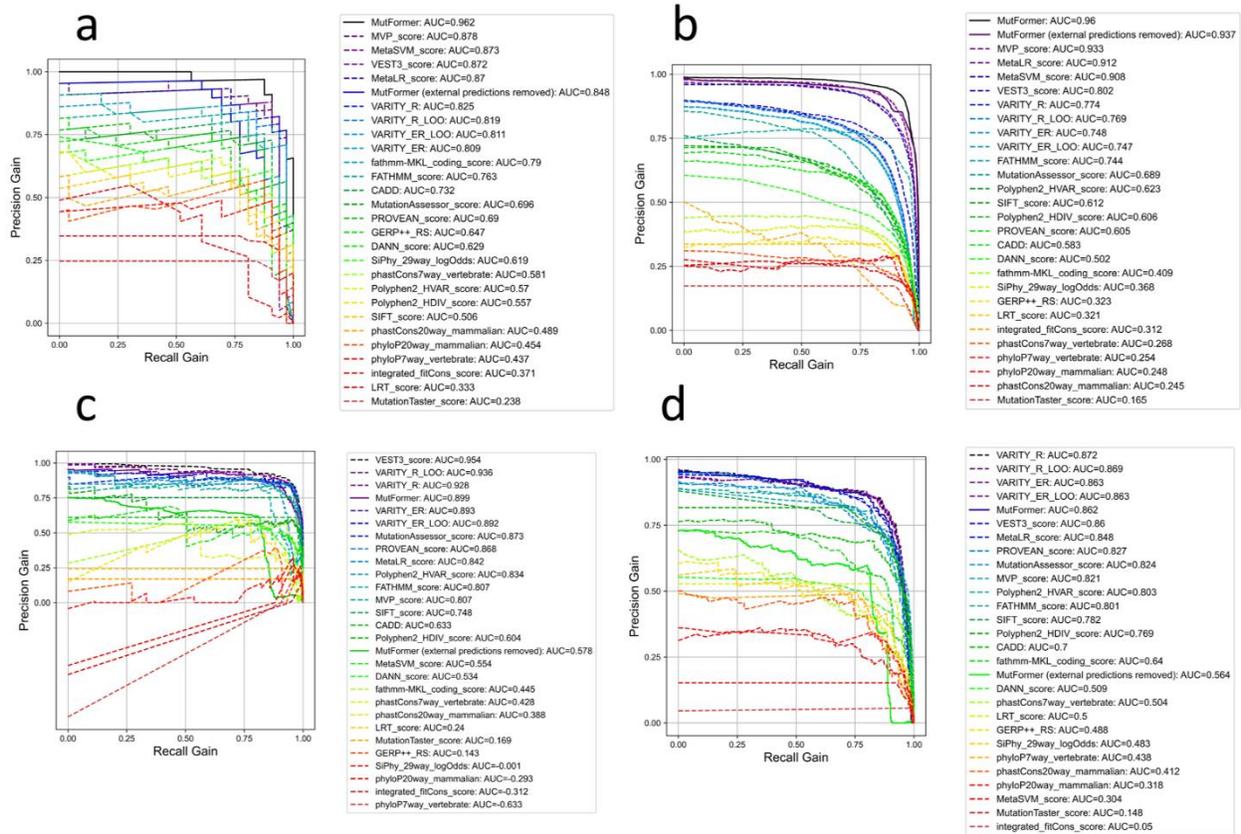

**Figure 5**